\documentclass[12pt,preprint]{aastex}

\shorttitle{DK Lac Low State}
\shortauthors{Honeycutt et al.}

\begin{document}


\title{The 2001-2003 Low State of Nova Lacertae 1950 (DK Lac)}


\author{R.K. Honeycutt\altaffilmark{1}, S. Kafka\altaffilmark{2}, 
H. Jacobson\altaffilmark{1,3}, A.A. Henden\altaffilmark{4},
D. Hoffman\altaffilmark{1,5}, T. Maxwell\altaffilmark{1,6},
J.W. Robertson\altaffilmark{7}, K. Croxall\altaffilmark{1,8}}

\altaffiltext{1}{Astronomy Department, Indiana University, Swain Hall West,
 Bloomington, IN 47405. E-mail: honey@astro.indiana.edu}

\altaffiltext{2}{Dept.of Terrestrial Magnetism, Carnegie Inst. of Washington,
 5241 Broad Branch Road NW, Washington, DC 2001. E-mail: skafka@dtm.ciw.edu}

\altaffiltext{3}{Current address: Michigan State University, Dept. of Physics 
 \& Astronomy, East Lansing, MI 48824-4540. E-mail: jacob189@msu.edu}

\altaffiltext{4}{American Association of Variable Star Observers, 49 Bay State Rd.,
 Cambridge, MA 02138-1203. E-mail: arne@aavso.org}

\altaffiltext{5}{Current Address: Astronomy Department, New Mexico State 
 University, Box 30001, Las Cruces, NM 88030-8001. E-mail: dhoffman@nmsu.edu}

\altaffiltext{6}{E-mail: tmaxwell@astro.indiana.edu}

\altaffiltext{7}{Arkansas Tech University, Dept. of Physical Sciences,
 1701 N. Boulder, Russellville, AR 72801-2222. E-mail: Jeff.Robertson@atu.edu}

\altaffiltext{8}{Current Address: Dept. of Physics \& Astronomy, University of 
 Toledo, Toledo, OH 43606.  E-mail: kevin.croxall@utoledo.edu}

\begin{abstract}

We report on extensive photometry of DK Lac obtained during the interval 1990-2009, 
which includes a 2 mag low state during 2001-2003.  Much of the photometry consists 
of exposures obtained with a typical spacing of several days, but also includes
26 sequences of continuous photometry each lasting 2 to 7 hours.  We find no 
evidence for periodicities in our data.  We do find that the random variations in 
the low state are $\sim$2$\times$ those in the high state, when expressed in 
magnitudes.   The lack 
of orbital-time-scale variations is attributed to the nearly face-on presentation 
of the disk.  There is a 0.2 mag decline in the high state brightness of the system 
over 19 years, which is consistent with the behavior of other old novae in the 
decades following outburst.  High-state spectra are also presented 
and discussed.  We find that the equivalent width of H$\alpha$ falls by 
$\sim$2$\times$ from 1991 to 2008.  The photometric properties are discussed in the 
context of the hibernation scenario for the behavior of novae between outbursts, 
in which we conclude that low states in old novae are probably unrelated to 
their possible entrance into hibernation. 

\end{abstract}


\section{Introduction}

DK Lac was a moderately fast (t$_3$ = 32 d ) nova in 1950, reaching V = 6 mag.
The nova light curve developed small eruptions midway down the decline
(e.g. Ribbe 1951), and the nebulae was spatially resolved (barely) by
Cohen (1985) and by Slavin, O'Brien \& Dunlop (1995); otherwise DK Lac
has rather undistinguished nova properties (Duerbeck 1981), and is not well-studied
in its post-nova stage.  As far as we are aware there has not been an orbital period 
study (spectroscopic or photometric) for DK Lac.

Nova-like (NL) cataclysmic variables (CVs) sometimes fade by 1-4 magnitudes,
remaining low from months to years; these are known as VY Scl stars.  The
VY Scl phenomenon is thought to be due to a 
cessation or diminution of the mass transfer from the secondary star to the
white dwarf, temporarily robbing the system of its accretion luminosity.  
In 2001 the autonomous 0.41-m
telescope at Indiana University found that DK Lac had entered a low
state (Henden, Freeland \& Honeycutt 2001) which lasted about
3 years.  A preliminary report on this low state, based on partial data
available at that time, appeared in Hoffman et al. 2003).   This current paper 
reports on nearly 20 years of sampled DK Lac 
photometry  surrounding this low state, plus continuous photometric 
sequences on numerous nights during both the low state and the subsequent 
return to the high state.  We also report on spectra acquired
during the high state. 
The DK Lac low state studied here is of interest because it is one of the
very few old nova to have displayed VY Scl-like behavior.  The character of the
low state therefore bears on how novae behave between nova outbursts and
whether they have long intervals of hibernation (Shara 1989) between nova eruptions,
without accretion luminosity.
  
\section{Data Acquisition and Reduction}

\subsection{Photometry}

Our DK Lac photometry is of two types.  First, we have long-term monitoring
at a typical cadence of 2-3 observations per week, obtained mostly with
the local Indiana automated photometric telescopes.  This program, operating
nearly continuously since 1990, is mostly used to monitor $\sim$120 NL
and old nova CVs for low states, both for what can be learned from the
photometry itself (e.g. Honeycutt, Cannizzo \& Robertson 1994; Honeycutt \& 
Kafka 2004; Kafka \& Honeycutt 2005), and to trigger low state spectrographic 
studies (e.g. Kafka et al. 2005a; Kafka, Honeycutt \& Howell 2006; Kafka et al. 
2008), as well as trigger more intensive low-state photometry (e.g. Kafka et al. 
2005a; 2005b; 2007; 2008).  To ensure good time coverage the Indiana 
telescopes are fully automated for autonomous operation, including
open-up/close-down decisions, dynamic scheduling, liquid nitrogen autofills,
acquisition of daily bias and flats, and data reduction.  No comprehensive
desscription of these systems was ever published, but much of the technical
details (Honeycutt et al. 1990; Honeycutt 1992; Honeycutt \& Turner 1992;
Honeycutt et al. 1994a) as well as motivations and strategies (Honeycutt 1994;
Honeycutt et al. 1994b; Honeycutt, Cannizzo \& Robertson 1994; Robertson,
Honeycutt \& Turner 1995; Honeycutt \& Kafka 2004; Kafka \& Honeycutt 2005)
are available.

The second variety of our DK Lac photometry consists of continuous monitoring
in sequences lasting 2-7 hours each.  These data were mostly acquired in
attended mode using a variety of telescopes and observers.   The motivation for
the monitoring was two-fold:  1) it was hoped that the orbital period might
be revealed, and 2) the low-state photometric behavior of VY Scl stars on time
scales of minutes to tens of minutes is not well defined, with several
interesting effects to be explored.  For example, in the low state of some 
CVs one sometimes sees 0.6 mag flares lasting 10-20 min, which have been
attributed to either stellar flares on the M dwarf doner star, or mass transfer
bursts (e.g. Honeycutt \& Kafka 2004 and references therein).

Table 1 is a log of the photometric observations.  Column 1 is a sequence 
designation (provided only for continuous exposures on a given night).  
Column 2 gives the date (or the date range), while column 3 provides
similar information for the JD.  Column 4 is the telescope, column 5 the 
filter, column 6 the exposure time in sec, column 7 the number of useable 
exposures, and column 8 the duration of the data stream.

The long-term photometry was acquired and processed in two different Indiana 
University (IU) observing programs.  The first program is an unattended, autonomous 
0.41-m telescope in central Indiana, informally called RoboScope.  This telescope 
collected V-band exposures of DK Lac 
for about 14 years 1990-2004 (the first entry in Table 1).  Flats and other 
detector calibration data were 
automatically acquired and applied after each exposurre, followed by aperture 
photometry and 
field identification, all using custom software (Honeycutt \& Turner 1992).  Final 
photometric reductions were done
using the incomplete ensemble technique contained in Astrovar, which is a custom 
package based on the technique described in Honeycutt (1992), but with the 
addition of a graphical user interface.  The second program for the long-term
photometry used an unattended, autonomous 1.25-m telescope at the same site as
RoboScope (the second entry in Table 2).  DK Lac exposures were acquired by this 
telescope for nearly 3 years 
2006-2009.  The real-time data reductions used for the RoboScope data were tightly
integrated into our old VMS operating system, while the 1.25-m telescope is
Linux-based.  The 1.25-m data were therefore reduced using a batch pipeline 
(not real-time) consisting of 
IRAF\footnote{IRAF is distributed by the National Optical Astronomy
  Observatories, which are operated by the Association of Universities for
  Research in Astronomy, Inc., under cooperative agreement with the National
  Science Foundation.} 
routines for detector calibrations, followed by the application of 
SExtractor\footnote{SEextractor is a source detection and photomery package
  described by Bertin and Arnouts 1996.  It is available from 
  http://terapix.iap.fr/soft/sextractor/.}.  The light curves were then generated 
using Astrovar.  The IU 1.25-m exposures of DK Lac were acquired during commissioning 
of the instrumentation.  The nightly dome flats gave poor results and we did
not have twilight flats on most nights.  Therefore we used median sky flats 
constructed from all the exposures in a given filter (typically 25-75) on a
given (or adjacent) night.  We had 
significant dark current from the
thermoelectically-cooled CCD that was in service on the IU 1.25-m at this time, 
which somewhat degraded the S/N.  The errors for the 1.25-m DK Lac data are 
typically 0.025 mag, about 2$\times$ the expected error if we had been able to
use high S/N dome flats and if the dark current were neglibible.  The IU 1.25-m 
exposures of DK Lac began with the V filter, but later
(after JD 2454330) switched to a Clear (C) filter.  However, we have 
no secondary standards for the C filter so the V standards were used instead.  This 
has little effect on the differential light curve, but does introduce some 
additional uncertainly into the zeropoint.

The DK Lac exposures which consist of continuous (or nearly continuous) sequences
on a given night (as opposed to the much more widely spaced automated long-term
photometry) originated from 3 different observing programs.  First we have
V-band sequences on 3 nights in 2001 from the 1.0-m telescope at the U.S. Naval
Obervatory, Flagstaff Station (USNO/FS) (entries 01-1 to 02-1 in Table 1).  These 
images were reduced using IRAF 
routines for both detector calibrations and aperture photometry.  Next we have 
V-band sequences on 15 different nights 2001-2006 using the 
0.91-m WIYN\footnote{The WIYN Observatory is 
  a joint facility of the University of Wisconsin-Madison, Indiana University, 
  Yale University, and the National Optical Astronomy Observatory} telescope at Kitt 
Peak (entries 03-1 to 06-5 in Table 1). Most of these images were reduced using 
IRAF for both detector calibrations 
and aperture photometry. However a few sequences used Cmuniwin\footnote{
  http://integral.sci.muni.cz/cmunipack/index.html}, a PC (Windows)-based 
photometry 
package orginally developed by Horch (1998).  These 0.91-m WIYN sequences were 
often 
(but not always) implemented as back-up programs for use on non-photometric nights,
and therefore sometimes have gaps due to passing clouds.  Finally we have sequences
on 8 nights 2007-2008 obtained with the IU 1.25-m telescope (entries 07-1 to 08-3
in Table 1), using the same
reduction pipeline described earlier for the long-term photometry of DK Lac on 
this telescope.  

We also have occasional DK Lac exposures (not parts of sequences) from the USNO/FS 
1.0-m telescope, from the WIYN 0.91-m telescope, and from 
the Tenagra Observatory\footnote{http://www.tenagraobservatories.com/}  
0.8-m telescope in southern Arizona; these occasional measures have been incorporated 
into our long-term light curve. 

The last reduction stage for all the DK Lac data used Astrovar, which provides 
photometric errors based on the repeatability of non-variable
stars as a function of instrumental magnitude.  These errors are for differential 
magnitudes with respect to the ensemble, which for DK Lac consisted of
86 nearby stars.  All of the DK Lac photometry 
used the secondary standards found in Henden \& Honeycutt (1997) to establish the
zeropoint.  The error in this zeropoint is typically $\sim$0.015 mag.  
The RoboScope data and the USNO/FS data have been
transformed to the standard UBV system using transformation coefficients evaluated
from regular B,V observations of standard star fields.  In applying this transformation
we assumed a color for DK Lac of 0.0 because B measures were usually not 
available for DK Lac; this approximation will introduce some additional error into 
the zeropoint.  
Finally, no transformations were applied for the WIYN 0.91-m data or for the IU 1.25-m
data (neither V nor C), resulting in additional zeropoint error for that data.  Overall
we estimate that the zeropoint error is $\lesssim$0.05 mag for all the data, and
significantly less for most of the data. 

DK Lac has a faint companion $\sim$5$\arcsec$ east.  On a low state USNO/FS
1.0-m image acquired 2001-Dec-17 UT (JD = 2452260.70278) in good seeing these two 
stars were measured  separately using DAOPhot PSF photometry in IRAF, with the 
following results: DK Lac V = 19.36$\pm$0.03, B-V = -0.17$\pm$0.03; Companion 
V = 19.56$\pm$0.03, B-V = 1.40$\pm$0.09.  In combined light, the companion 
contributes $\sim$30-50\% of the V-band low-state light (depending on the variable 
low-state brightness of DK Lac), and $\sim$9\% of the high-state light.
It was not practical to do PSF photometry on all our images because the
two stars are not resolved in poor seeing nor when the system is in the high state.  
Therefore we chose to use a numerical 
diaphragm for DK Lac that includes both stars, and correct the resulting
magnitude for the contribution of the companion.  This means that our high-state
magnitudes are $\sim$0.10 mag fainter than the magnitudes which other observers
might report for the same high-state epoch, if they are unaware of the contribution \
from the companion.  

Figure 1 shows our full light curve (with 2304 points plus 92 limits), where we have 
plotted the magnitudes corrected for the light of the companion.  The error bars 
(which are not shown in Figure 1, but which are present on subsequent light curve
plots) are for differential magnitudes with respect to the ensemble.  Table 2 lists 
all the magnitudes and errors (for the differential magnitudes).  The complete 
version of Table 2 is available only in electronic form.  The data will also be 
archived with the AAVSO.

In the low state DK Lac is mostly just beyond the magnitude limit for RoboScope, 
resulting in numerous upper limits (which nevertheless help define the duration
of the low state) in Figure 1.  (When a limit is obtained, this means that the 
combined light of DK Lac and the companion was not detected.  Technically then, 
the limits are for a different  circumstance than for the magnitudes, which are 
for DK Lac only, and the limits are actually fainter
(by an unknown amount) than shown).  Labels 01-08 in Figure 1 designate continuous 
sequences; see Table 1 and the caption to Figure 1 for further information.

It is seen in Figure 1 that DK Lac experienced a 2 mag deep low state beginning
2001-Jan and ending sometime between 2003-Jul and 2004-Jun, for a duration
of 2.5-3.4 years.  There also appears to be an overall decline of $\sim$0.2 mag in the
high state brightness over our 19 years of data.

\subsection{Spectroscopy}

Spectra of highly magnetic NL CVs (polars) in the low state typically have
emission lines arising from the inner hemisphere of the doner star.  These
lines are not single-peaked but have satellite components whose origin remains
uncertain (Kafka et al. 2005a, 2005b, 2007; 2008).  These satellites may
arise from large-scale magnetic structures on the secondary star, or magnetic 
structures connecting the secondary star to the white dwarf.  A central question 
is whether this
behavior is induced by the strong magnetic field of the white dwarf or is
a property of "hyperactivity" on the secondary star.  It is therefore important
to obtain low-state spectra of CVs which are disk systems (such as DK Lac) 
rather than polars, in order to help distinguish these two possibilities. 

Our attempts at low-state spectroscopy of DK Lac were unsuccessful because of
system faintness.  By the time we were able to 
obtain spectra using larger telescopes, DK Lac had returned to the high state.  
Figure 2 is an expanded portion of the DK Lac light curve in Figure 1, with the times of 
our spectra marked.   Table 3 is a journal of our spectroscopic observations.

Spectral sets S1 and S4 were obtained using the GoldCam slit spectrograph on the 
KPNO\footnote{Kitt Peak National Observatory is a division of the National 
  Optical Astronomy Observatory, which is operated by the Association of Universities 
  for Research in Astronomy, Inc., under cooperative agreement with the National 
  Science Foundation} 2.1-m telescope.  Grating 35 was used in first order, 
providing coverage $\sim$5000-8100\AA\ at $\sim$3\AA\ resolution. 
Spectral set S2 was obtained using the MOS/Hydra multiple object spectrograph 
on the WIYN telescope.  Grating 600 was used in first order, providing
coverage $\sim$5300-8200\AA.  The ``red'' 2$\arcsec$ fiber bundle was employed,
yielding $\sim$3\AA\ resolution; numerous other fibers were used for sky
subtraction.  Spectral set S3 used the 6.5-m MMT\footnote{The MMT Obervatory is a 
 joint facility of the Smithsonian Institution and the University of Arizona} 
at Mt. Hopkins, Arizona.  The Blue Channel slit spectrograph was employed with a 
500 line/mm grating, providing coverage $\sim$5000-8100\AA\ at a resolution of 
$\sim$3.6\AA.  
  
For all spectra our detector calibrations used standard IRAF
procedures, and for spectral extractions and wavelength calibrations we used 
IRAF's onedspec/twodspec packages.  No spectrophotometric calibrations were
applied for any of the spectra, and the continua in the reduced spectra were 
nomalized to unity.  Also, no corrections were made for telluric spectral 
features  in the near-IR.  Each spectral sequence contained 2-4 exposures 
(see Table 2), which were combined after reductions.  

\section{Analysis}

\subsection{Photometry}

Figures 3 and 4 are representative plots of the light curves of the individual 
photometric sequences listed in Table 1.  Figure 3 contains two
nights of low-state data, demonstrating that some nights have little variation,
while other have systematic 1 mag changes over several hours.  Furthermore the
mean low-state mag can change by 0.5 mag between nights.   Figure 4 is an example of
a night of high-state photometry, where we see flickering-type behavior with
little change in mean mag over 7 hours. 

The location of the sequences in the long-term light curve are labeled in Figure 1.  
We discuss separately below the low-state and the high-state photometry.  

\subsubsection{The Low State}

Livio \& Pringle (1994) proposed that VY Scl low states are due to a large starspot
drifting into the L1 region of the doner star.  By analogy to sunspots (which are 
observed to have signifcantly lower gas densities than the surrounding photosphere)
it was argued that a starspot at L1 will greatly lower $\dot{M}$, reducing the accretion
luminosity and producing a VY Scl low state.  VY Scl stars typically take a few
tens of days to enter a low state.  In the starspot hypothesis the ingress and egress 
time scales of low states are not well-contrained because they depend on the 
poorly-known drift rate of star spots on CV doner stars.  Honeycutt \& Kafka (2004) 
reported that the shapes of many VY Scl ingress and egress events were dual-sloped,
always being faster when fainter.  It was argued that this effect is consistent with
the starspot scenario in a number of respects, if the two slopes are associated with
the passage of the umbral and penumbral portions of the starspot drifting across L1.      

Figure 2 shows a few data points on the decline of DK Lac into the low state, especially 
the early  stages.   However later portions of the decline have few points because 
RoboScope does not reliably reach fainter than V$\sim$18. The recovery from the low state 
was almost totally missed.  The low state had an amplitude of 2.0 mag and (taking into 
account the missing-data intervals) lasted between 2.5 and 3.4 years.  Figure 5 shows an 
expanded portion of Figure 1 where we have characterized the decline as two straight 
lines by appealing to the behavior of better-sampled high-state low-state transitions 
of VY Scl stars in Honeycutt \& Kafka (2004).  In that
work it was found that, for $\sim$40\% of the observed transitions, a pronounced 
slope change occurred during the transition to and from the low state.  That slope 
change was always in the sense that the transition was faster when fainter.  The 
speed of the transitions was characterized by an e-folding time, $\tau$.  For 17 
transitions  having dual slopes in  Honeycutt \& Kafka (2004), it was found that 
$\tau_{faint}$ ranged from 3 to 40 days (with a mean of 15 days), while 
$\tau_{bright}$ ranged from 4 to 160 days (with a mean of 52 days).  For
the straight lines plotted in Figure 5 we have for DK Lac $\tau_{faint}$ = 12 days 
and $\tau_{bright}$ = 175 days, values within or near the range for other VY Scl stars.

The photometric sequences acquired during the low state fall naturally into 
two sets.  Set 1 consists of 3 sequences obtained over a 21 night interval between 
2001-Dec-18 and 2002-Jan-07 (Sequences 01-1 to 02-1 in Table 1), while Set 2 
consists of 7 sequences obtained 20 months later, over an  
18 night interval in 2003-Jun/Jul (sequences 03-1 to 03-7).  
Figure 6 shows the mean magnitudes for the sequences from 2001-02 (Set 1) alongside
a similar plot for the sequences from 2003 (Set 2).  There does not appear to be any
systematic change in the mean low-state brightness over these intervals.  The short 
error bars (with caps) are for the standard deviation of the mean (sdm), showing that 
night-to-night variations substantially exceed the sdm of the individual sequences.  
The long error bars (without caps) are for the standard deviation of a single 
observation (sdso), which should be independent of the number of points in the 
sequence, measuring instead how ``noisy'' or ``quiet'' is each photometeric sequence.
There is no apparent systematic change in sdso as the low state progresses, nor any 
correlation of sdso with low-state mean magnitude.  The mean value of sdso for the 
10 low-state sequences is 0.16 magnitude, and the mean low-state magnitude is 19.0.      

Our data provide little leverage to characterize the low-state night-to-night 
variations.  Therefore we searched for periodicities in the expected orbital period
range of 1 to 12 hours, using data which we pre-whitened by adjusting the level 
of each sequence to a common mean magnitude.  Periodograms (not shown) were constructed 
for a variety of groupings of the low-state data.  
In these power spectra there are no isolated peaks that stand out well from the noise and 
the alising background.  Instead we see that the major power is distributed primarily 
between 2 and 3 hours for the 2001-2002 data, but primarily between 3 and 5 hours 
for the 2003 data.  We folded the data on the highest peaks, but the resulting light curves
did not provide any additional insight into the nature of the variations.  When we 
constructed new periodograms by randomizing the magnitudes, 
keeping the same JDs, it was concluded that the excess power at these characteristic 
frequencies is real, as is the change in the characteristic time scale.  

\subsubsection{The High State}

Periodograms of the high-state photometry were examined using pre-whitened data and a variety
of data groupings.  For the continuous sequences we find broad power from 2 to 3.5 hours, with
somewhat isolated peaks at 0.11162(5) and 0.08431(3) day for some data groupings..  The light 
curves folded on these two periods have sinusoidal amplitudes of $\sim$0.03 mag with superimposed 
scatter of $\sim$0.2 mag.  It seems likely that all these "periods" are spurious, but we mention them
for completeness.

The IU 0.41-m data of 1990-2002 consists of single exposures acquired at typical intervals of a 
few days.  This cadence could hardly be more 
different than that for the sequenced photometry, which typically consists of 20-80 exposures
over a few hours of a given night (see Table 1).  The nights containing sequences are typical of
conventionally-scheduled observing runs, consisting of widely-separated sets of several
adjacent or nearly-adjacent nights.  The combination of these two cadences would seem to be ideal
for establishing any reliable periodicities or quasi-periodicities, especially since the
errors for the sequences and for the sampled photometry are similar.  However, there are no
significant peaks in common between the periodograms of the sequenced photometry and the sampled 
photometry,
suggesting that no stable periodicities are present in the high-state data.  Even the concentrated
power between 2 and 3.5 hours seen in the periodograms of the sequence photometry is largely missing
in the periodogram of the sampled photometry.  

With regard to the B-V color of DK Lac in the high state, Szkody (1994) reported 0.08 and 
Ringwald et al. (1996) gave 0.1.  Ringwald et al. also provide E$_{B-V}$ = 0.45, which implies
an intrinsic color corresponding to a blackbody with T$\gtrsim$50,0000$\arcdeg$.  

As for the high-state spectra, Figures 7, 8, 9, and 10 show the combined spectra for 
sets S1-S4.  The best S/N 
is found in S3 (Figure 9), although it has some residual near-IR fringing longward 
of H$\alpha$.  In S3 we see in emission H$\alpha$, HeI 5876, and HeI 6676.  
HeI 6676 is also visible in S2 (Fig 8) and S4 (Fig 10).  The NaD  doublet 
is in absorption with a total equivalent width of 1.8\AA.

\section{Discussion}

Distance, extinction, and M$_V$ are not well known for DK Lac.
Duerbeck (1981), using a calibration of light curve decay time with luminosity (along with
an adopted extinction of 1.2 mag) estimates a distance of 1500$\pm$200 pc, while Slavin et al. (1995) 
used the method of nebular expansion parallax to derive 3900$\pm$500 pc.  In our spectrum S3 the
EW of Na~D1 is 0.83\AA.  Using the calibration found in Munari and Zwitter (1997) we find
A$_V$ = 1-3 mag.  This wide range is due to saturation effects along with uncertainties
introduced by unresolved multiple line components.  Using the nebular expansion parallax
distance and adopting A$_V$ = 1.2$\pm$0.2 we find M$_{V,max}$ = -9.1$\pm$0.4, and 
M$_V$ = 2.9$\pm$0.4 59 years after the outburst, in 2009.  These M$_{V}$ values are
consistent with results for other novae  at maximum and for old novae decades
after the nova (Warner 1995).

The extinction is unlikely to be
much higher than our adopted value because then the intrinsic color becomes too blue for
a blackbody, requiring the assumption of strong emission lines or other non-thermal contributions
to the B-V color.  Interestingly, the color of DK Lac assumed an even bluer value of -0.17 in the
low state.  This result is difficult to understand.  Most CVs become redder in the optical
when accretion luminoisity falls, as the cool secondary star begins to dominate.  Perhaps the
secondary star in DK Lac is of particularly low luminosity and/or the accretion luminosity
did not dissappear completely.  This blue color in the low state is based on only a single
measurement, but nevertheless appears secure (see $\S$ 2.1).

The inclination of the system can
be estimated using the calibrations found in Warner (1986).  Using the correlation of
H$\alpha$ EW with orbital inclination, we find that i = 0-50$\arcdeg$ for DK Lac.  The correlation
of M$_V$ with disk inclination is more confining, yielding i = 0-25$\arcdeg$.  It seems
clear that the disk of DK Lac is presented nearly face-on.  This may account for the
lack of any orbital signature in the photometry reported in this paper.  The effect of various 
kinds of non-axisymmetic accretion structure(s) on the light curve is minimized when the
portions of the disk seen by the observer are always in view, as is the case at low
orbital inclination.

In fact both the high-state and low-state photometry appear random in nature, though
perhaps with preferred time scales.  For the sampled high-state photometry obtained prior 
to the low state, prewhitened by observing season, the rms scatter is 0.094 mag.  For the
sequenced high state photometry after the low state, prewhitened by sequence, the rms
scatter is 0.085 mag.  Let us take 0.09 mag rms as the characteristic scatter (flickering), 
during the high state, to be compared to 0.16 mag rms in the low state.

VY Scl stars do not often stay in the low state long enough to characterize their low
state variations, but when they do (e.g. MV Lyr in Honeycutt \& Kafka 2004) the light curve
seems flat to within $\sim$0.15 mag, with brief 0.5 mag flares superimposed.  This behavior
is similar to that seen in polars, where the low states are more frequent (see low-state
light curves of AM Her, ST LMi, and AR UMa in Kafka \& Honeycutt 2005).  Schmidtke et al.
(2002) reports on sequenced photometry obtained as LQ Peg recovered from a low state,
finding 0.09 mag variations, with no coherent signal between 1 and 6 hours; these
sequences were obtained roughly mid-way up the recovery to the high state.  Schmidtke et al.
(2002) found that as LQ Peg brightened from 15.8 to 14.5 the flickering decreased from
0.09 to 0.02 mag.  This is in the same sense as the change we find in DK Lac.  However, as
DK Lac brightened from 19.0 to 17.0, the amount of flickering declined from 0.16 to 0.09 mag,
a less extreme change than seen in LQ Peg.  Schmidke et al. also found that the flickering
was a constant fraction of the intensity as the system recovered from the low state.
That is clearly not the case in DK Lac, in which the high state rms variation expressed in 
intensity units is 3-4$\times$ the low state rms variation in intensity units. 

The hibernation scenario for cyclic nova episodes (see Shara 1989 for a review) holds that
at some point after the nova outburst accretion drops to a near zero, a state known
as hibernation; the system is thought to remain in hibernation for most of the interval 
between nova outbursts.  Advantages of hibernation include reducing the mean accretion 
rate to levels
needed to allow the next thermonuclear runaway, and providing consistency between nova
rates and cataloged CVs because of the inconspicious nature of hibernating systems.
The scenario has received criticisms (e.g. Naylor et al. 1992) but remains a topic of
considerable interest (e.g. Martin \& Tout 2005).  The fact that post-nova systems
fade by 1-2 mag per century following the nova (Vogt 1990; Duerbeck 1992) is consistent
with the hibernation scenario, but it may also be the case that the onset of low states
marks the transition into hibernation. 

At least one other old novae has had a low state subsequent to the nova outburst.
V533 Her (Nova Her 1963) experienced two prominent low states in 1995-June and 1996-Jan, each lasting
2-3 months, with depths of 1.5 mag (Honeycutt \& Kafka 2004).   These V533 Her
low states were too brief for in-depth study.  The DK Lac and V533 Her low states occurred
51 years and 32 years respectively after the nova.  Neither system was monitored regularly
prior to 1990, so other low states might have been missed.  Honeycutt \& Kafka found that
in 12 years of monitoring 65 old novae and NL CVs, $\sim$15\% showed at least one low
state exceeding 1.5 mag.  Statistics for the old novae portion of the sample are not as good
as desired, but there is nevertheless no evidence that the rate of low states in old novae 
(2 of $\sim$24) is any
different from that for NLs, at least when the old novae are examined 30-60 years after the nova.
One might expect that if low states mark the manner in which old novae approach
hibernation, then low states would be more frequent and/or more prolonged as the old novae
became NLs.  Our tentative conclusion is that low states are unrelated to hibernation, as
supported  by the fact that the low state in Fig 1 seems to be photometrically independent
of the slow high-state decline, appearing to simply be superimposed on the light curve
as an independent phenomenon.  This is consistent with low states being due to a short-term
effect rather than part of an evolutionary progression, the migration of starspots under
the inner Lagragian point (Livio \& Pringle 1994) being a well-regarded mechanism.

Ignoring the low state interruption, the brightness of DK Lac is seen in Figure 1 to decline 
rather smoothly by 0.2 mag over 19 years.  Activity cycles on the secondary stars of CVs
have long been suspected to modulate the mass transfer rate and hence change the 
brightness of CVs on time scales of decades (e.g. Warner 1988; Bianchini 1990; Richman, 
Applegate \& Patterson 1994;
Ak et al. 2001).  However, we do not think that this effect is responsible for the
0.2 mag decline, for two reasons.  First, there is no inflection in the decline as might
be expected for cyclic behavior shorter than 3 decades.  Second, the decline rate
is consistent with the rate found for other old novae many decades after outburst.
For nine old novae Duerbeck (1992) finds a mean rate of decline 50 years after outburst 
of 10$\pm$3 millimag per year.  This is attributed to the slow decline of irradiation-induced 
mass transfer, and is concluded to be consistent with the hibernation hypothesis.   
The rate in DK Lac is 10.5 millimag per year, fully consistent with the behavior of other 
post-novae.  

Over roughly the same interval in which the continuum of DK Lac declines by 0.2 mag, the EW of 
H$\alpha$ declines by $\sim$2$\times$. Because the continuum is falling at the same time, 
this weakening of H$\alpha$ is actually even more pronounced.
In Figure 11 we plot the EW of H$\alpha$ vs JD, where we have added 
a 1991 data point from Ringwald et al. (1996).  The suggestion of a linear decline
in H$\alpha$ EW over 16 years is intriging, but needs confirmation.   The initial fall in
$\dot{M}$ (and therefore in system brightness) following the nova eruption 
is thought to be due to declining irradiation of the doner star as the white dwarf cools
following the nova eruption (Duerbeck 1992 and references therein).
The hibernation scenario (Shara 1989) holds that $\dot{M}$
continues to fall even further, taking the system through the dwarf nova (DN) regime and finally into
a detached binary state with little or no accretion luminosity.  In general,
the EW of DN emission lines is greater than that of emission lines in old novae, opposite to the
trend in Figure 11.  Therefore we suggest that the emission lines in DK Lac arise
mostly from the illuminated secondary star rather than from the accretion disk.  The breadth of
the emission lines in DK Lac are small (FWHM$_{H\alpha}$ = 12\AA, FWHM$_{HeI}$ = 8\AA) compared
to most old nova.  While much of the narrowness can be attributed to the low system inclination,
it is also consistent with the emission arising from the irradiated inner hemisphere of the
secondary star. 

\section{Summary}

Long-term photometry of Nova Lac 1950 (DK Lac) from 1991 to 2008 shows a 2 magnitude deep 
VY Scl-type low state beginning 2001-Jan and lasting for between 2.5 and 3.4 years. 
The shape of the ingress into the low state is similar to that of other VY Scl stars,
having a more rapid fall as the ingress proceeds.  There is also an overall decline of
0.2 magnitude in the high state brightness over our 19 years of data, consistent with
the fading rate of other old novae when measured several decades after the nova eruption.

Our photometry also includes 26 continuous sequences each lasting 2 to 7 hours.  We searched
unsuccessfully for photometric periodicities in the range of 1-12 hours (the range of expected
orbital periods)
using both the long-term and the sequenced data, and both the high and low state data.
The lack of orbital-time-scale variations is attributed to the nearly face-on presentation 
of the disk.  High-state spectra are also presented 
and discussed.  We find that the equivalent width of H$\alpha$ falls by $\sim$2$\times$ from 
1991 to 2008.   

The fact that this low state occured in an old novae suggests that it might be connected to
the entrance of DK Lac into hibernation.  However no distinction is found between the low
states of old nova and those of NL CVs (which were presumably unrecorded novae hundreds
of years ago).  Instead we conclude that low states are unrelated to the possible entrance 
of a system into hibernation.

\acknowledgments

ACKNOWLEDGMENTS

Constantine Deliyannis, Caty Pilachowski, and their students, had programs on the WIYN 0.91-m 
that required photometric weather.  When the weather was mostly clear but non-photometric,
DK Lac frequently became the back-up target.

Some the initial work on this project was done by Doug Hoffman as a Research Experience
for Undergraduates (REU) student at Indiana University, supported by the National
Science Foundation.

\begin{deluxetable}{lllcccrc}
\tablenum{1}
\tablewidth{0pt}
\tablecolumns{8}
\tablecaption{Photometry Log}
\tablehead{\colhead{Seq.} & \colhead{UT} & \colhead{JD} & \colhead{Tel.} 
&\colhead{Filt.} & \colhead{Secs.} & \colhead{\# Exps.} &\colhead{Dur.}}
\startdata
      & 1990-Nov-12 to & 2448207 to  & IU 0.41-m     & V   & 240     & 739 & 14.1 yr  \\
      & 2004-Dec-04    & 2453343     &               &     &         &     &          \\
      &                &             &               &     &         &     &          \\
      & 2006-Aug-23 to & 2453947 to  & IU 1.25-m     & V,C & 150     & 251 & 2.8 yr   \\
      & 2009-May-21    & 2454972     &               &     &         &     &          \\
      &                &             &               &     &         &     &          \\
01-1  & 2001-Dec-18    & 2452261     & USNO/FS 1.0-m & V   & 120     &  54 & 3.4 hr   \\
01-2  & 2001-Dec-20    & 2452263     & USNO/FS 1.0-m & V   & 120     &  50 & 3.1 hr   \\     
02-1  & 2002-Jan-07    & 2452281     & USNO/FS 1.0-m & V   & 120     &  60 & 2.7 hr   \\
      &                &             &               &     &         &     &          \\     
03-1  & 2003-Jun-29    & 2452819     & WIYN 0.91-m   & V   & 90-150  &  55 & 3.3 hr   \\     
03-2  & 2003-Jun-30    & 2452820     & WIYN 0.91-m   & V   & 120-240 &  63 & 4.6 hr   \\     
03-3  & 2003-Jul-02    & 2452822     & WIYN 0.91-m   & V   & 100-500 &  35 & 4.1 hr   \\     
03-4  & 2003-Jul-03    & 2452823     & WIYN 0.91-m   & V   & 100     &  87 & 4.2 hr   \\     
03-5  & 2003-Jul-10    & 2452830     & WIYN 0.91-m   & V   & 180     &  40 & 3.0 hr   \\     
03-6  & 2003-Jul-11    & 2452831     & WIYN 0.91-m   & V   & 180     &  33 & 3.4 hr   \\     
03-7  & 2003-Jul-16    & 2452836     & WIYN 0.91-m   & V   & 180     &  26 & 2.8 hr   \\     
      &                &             &               &     &         &     &          \\     
04-1  & 2004-Jun-29    & 2453185     & WIY 0.91-m    & V   &  60     &  43 & 1.6 hr   \\     
04-2  & 2004-Jul-03    & 2453189     & WIY 0.91-m    & V   &  60     &  18 & 0.9 hr   \\     
04-3  & 2004-Jul-05    & 2453191     & WIY 0.91-m    & V   &  60     &  70 & 2.4 hr   \\     
      &                &             &               &     &         &     &          \\     
06-1  & 2006-Sep-30    & 2454008     & WIY 0.91-m    & V   & 120     &  35 & 2.0 hr   \\     
06-2  & 2006-Oct-01    & 2454009     & WIY 0.91-m    & V   & 120-300 & 109 & 7.0 hr   \\         
06-3  & 2006-Sep-03    & 2454011     & WIY 0.91-m    & V   & 120-300 &  77 & 7.0 hr   \\     
06-4  & 2006-Oct-04    & 2454012     & WIY 0.91-m    & V   & 120-300 &  18 & 1.3 hr   \\     
06-5  & 2006-Oct-05    & 2454013     & WIY 0.91-m    & V   & 120-300 &  40 & 4.0 hr   \\     
      &                &             &               &     &         &     &          \\     
07-1  & 2007-Aug-18    & 2454330     & IU 1.25-m     & V   & 180     &  70 & 4.6 hr   \\     
07-2  & 2007-Sep-04    & 2454347     & IU 1.25-m     & C   & 180     &  70 & 4.7 hr   \\     
07-3  & 2007-Sep-14    & 2454357     & IU 1.25-m     & C   & 180     &  68 & 4.6 hr   \\     
07-4  & 2007-Oct-07    & 2454380     & IU 1.25-m     & C   & 120     &  63 & 2.7 hr   \\     
07-5  & 2007-Oct-10    & 2454383     & IU 1.25-m     & C   & 120     &  65 & 3.9 hr   \\     
      &                &             &               &               &     &          \\     
08-1  & 2008-Oct-03    & 2454742     & IU 1.25-m     & C   & 150     &  46 & 4.5 hr   \\     
08-2  & 2008-Oct-04    & 2454743     & IU 1.25-m     & C   & 150     &  20 & 2.4 hr   \\     
08-3  & 2008-Oct-05    & 2454744     & IU 1.25-m     & C   & 150     &  41 & 4.3 hr   \\ 
\enddata
\end{deluxetable}
\clearpage

\begin{deluxetable}{cccc}
\tablenum{2}
\tablewidth{0pt}
\tablecolumns{4}
\tablecaption{Magnitudes for DK Lac}
\tablehead{\colhead{JD}  & \colhead{V} & \colhead{Error} & \colhead{Source}}
\startdata
2448207.68948  & 16.943  & 0.034  & RoboS \\
2448208.63269  & 16.841  & 0.031  & RoboS \\
2448208.76882  & 16.862  & 0.044  & RoboS \\
2448209.64547  & 16.938  & 0.031  & RoboS \\
2448209.73172  & 16.923  & 0.040  & RoboS \\
2448209.80370  & 16.899  & 0.048  & RoboS \\
2448233.72259  & 17.054  & 0.063  & RoboS \\
2448234.61401  & 16.919  & 0.044  & RoboS \\
2448234.71393  & 16.872  & 0.052  & RoboS \\
2448235.58032  & 16.907  & 0.039  & RoboS \\
.............  & ......  & .....  & ..... \\
.............  & ......  & .....  & ..... \\
.............  & ......  & .....  & ..... \\
2454972.82035  & 17.130  & 0.029  & IU 1.25-m \\
\enddata
\end{deluxetable}
\clearpage

\begin{deluxetable}{cccccc}
\tablenum{3}
\tablewidth{0pt}
\tablecolumns{6}
\tablecaption{Spectroscopy Log}
\tablehead{\colhead{Set}  & \colhead{UT} & \colhead{JD} & \colhead{Tel}
& \colhead{Secs} & \colhead{\# Exps}}
\startdata
S1 & 2005-Oct-12 & 2453655 & KPNO 2.1-m & 600-900 & 2 \\
S2 & 2005-Oct-26 & 2453669 & WIYN 3.5-m & 600     & 3 \\
S3 & 2006-Sep-16 & 2453994 & MMT  6.5-m & 900     & 2 \\
S4 & 2008-Jun-07 & 2454624 & KPNO 2.1-m & 900     & 4 \\
\enddata
\end{deluxetable}

\clearpage


\begin{figure}  
\epsscale{0.7}
\plotone{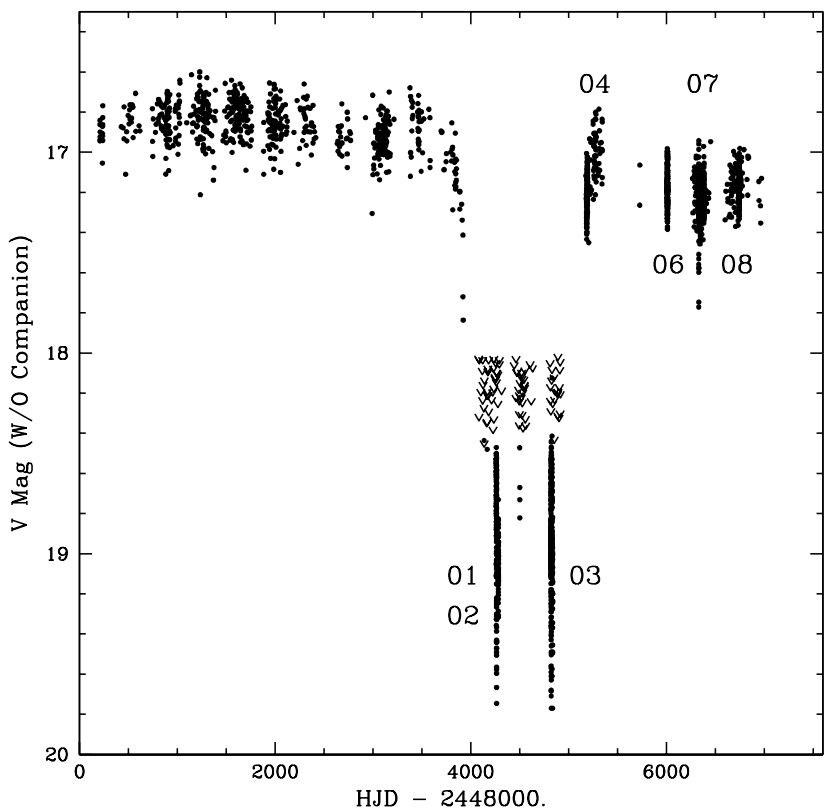}
\caption{The full light curve of DK Lac from 1990-Nov-12 to 2009-May-21.  The
contribution of the companion star has been removed (see text for details).  Error
bars have been omitted for clarity.  The upper limit symbols during the
2001-03 low state are from RoboScope.  The vertical clusters of unresolved
points labeled 01 through 08 are continuous sequences of exposures, typically
obtained on several successive or nearly successive nights.  The sequence root
number is the last two digits of the year.  A serial number is appended to the
sequence root to form the sequence number column in Table 1, and this notation is 
preserved for plots of the individual sequences (Figures 3 and 4)} 
\end{figure}
\clearpage  

\begin{figure}  
\epsscale{0.9}
\plotone{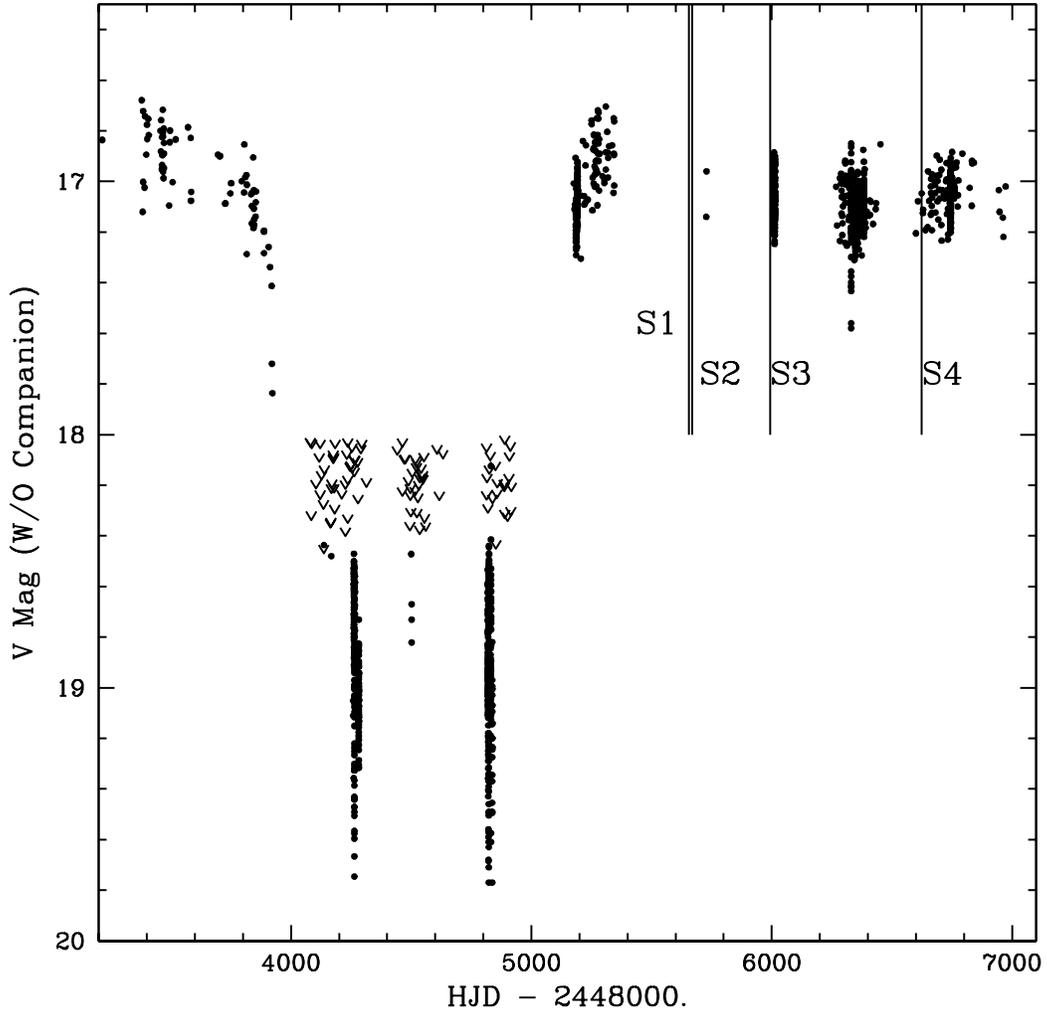}
\caption{ The latter portion of Figure 1 with the
times of spectroscopic data marked.  The labels for the times of
spectral exposures correspond to column 1 of Table 3.} 
\end{figure}
\clearpage

\begin{figure}  
\epsscale{0.85}
\plotone{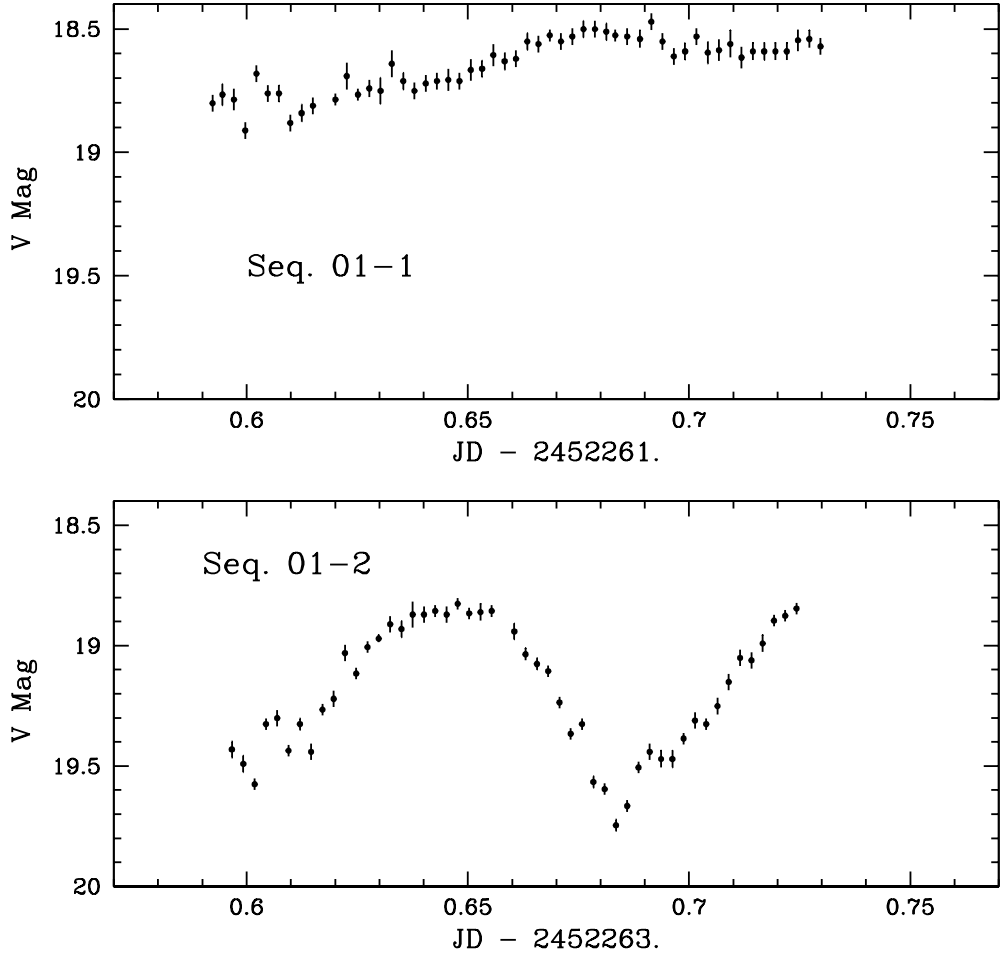}
\caption{Examples of low-state photometric sequences.  The 
sequence notation refers to labels in Figure 1 and Table 1.  The V magnitudes
have been corrected for the contribution of the close companion.
This plot is for 2 sequences obtained 2001-Dec-18(UT) and 2001-Dec-19(UT).
Error bars are plotted but are often too small to be seen.} 
\label{xx2}
\end{figure}
\clearpage  

\begin{figure}  
\epsscale{0.9}
\plotone{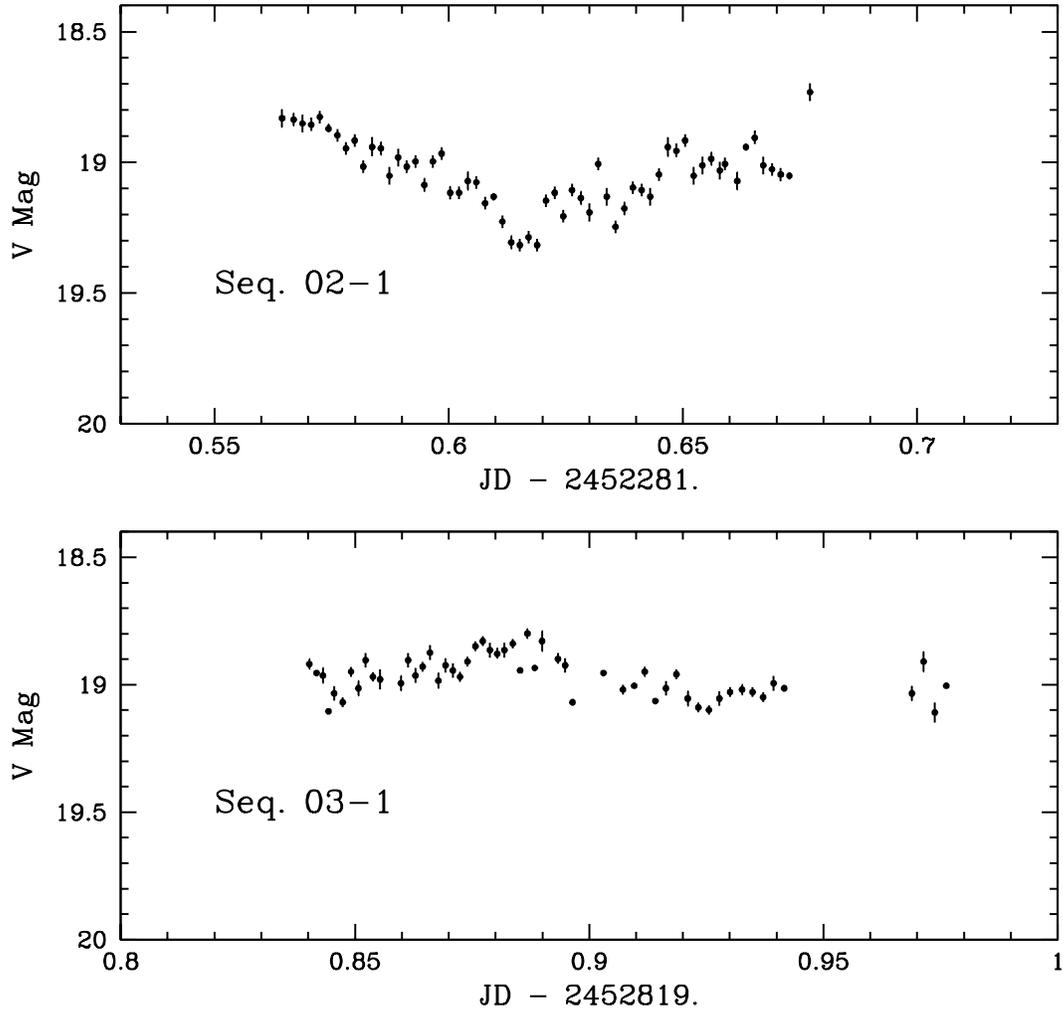}
\caption{An example of a high-state photometric sequence.  The 
sequence notation refers to labels in Figure 1 and Table 1.  
Note that the data in these two panels overlap somewhat.
The V magnitudes have been corrected for the contribution of the 
close companion. This plot is for a sequence obtained 2006-Oct-01(UT).}
\end{figure}
\clearpage  

\begin{figure}  
\epsscale{0.9}
\plotone{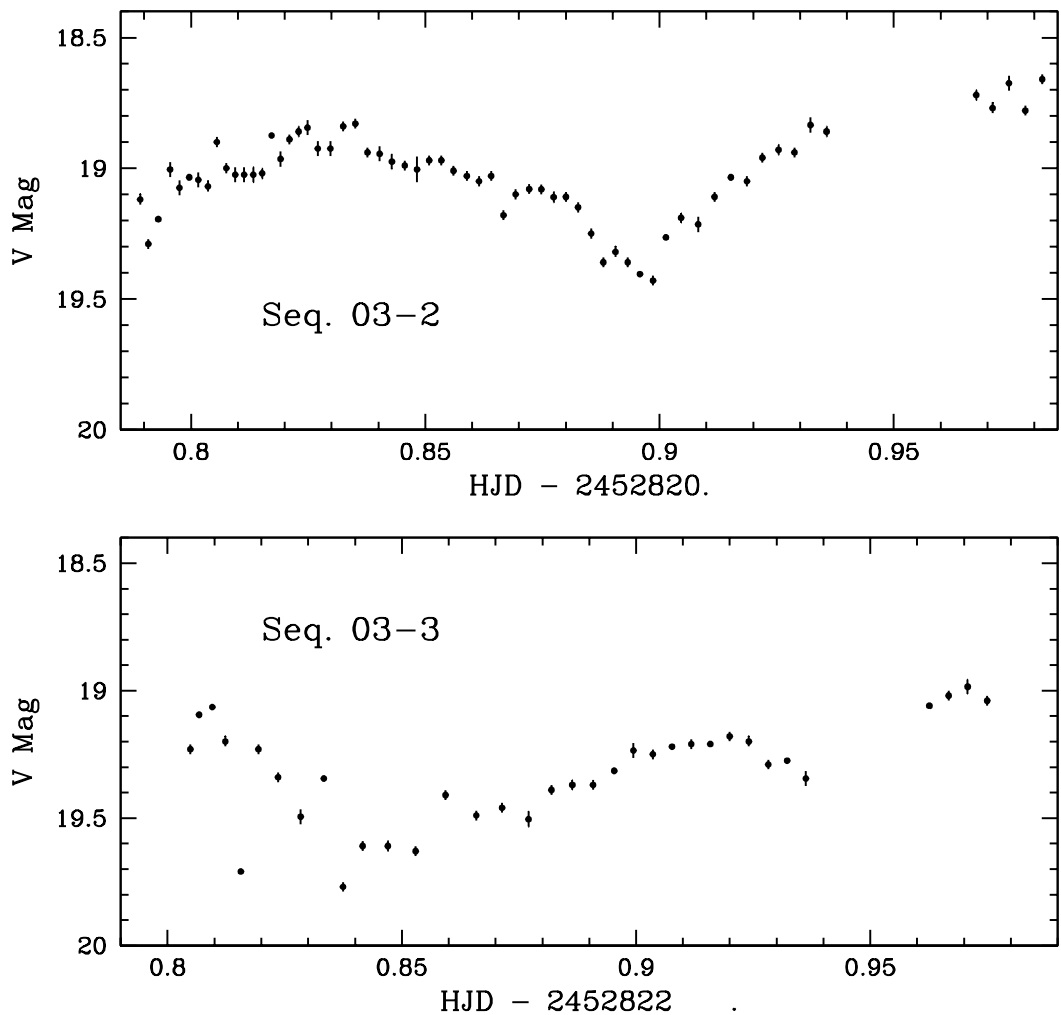}
\caption{An expanded view of the portion of Figure 1 containing the ingress
to the low state.  The characterization of the shape of the ingress as two straight
lines is adapted from the behavior of high-state/low-state transitions in numerous
other VY Scl stars, from Honeycutt \& Kafka 2004.  The level of the low state is
fixed at V = 19 using data off the plot to the right (see Figure 1), where the low
state magnitude varies between 18.5 and 19.5.  The occasional RoboS detections in
near 18.5 in the plot are presumably the peaks of this distribution of low-state
magnitudes.} 
\end{figure}
\clearpage  

\begin{figure}  
\epsscale{0.9}
\plotone{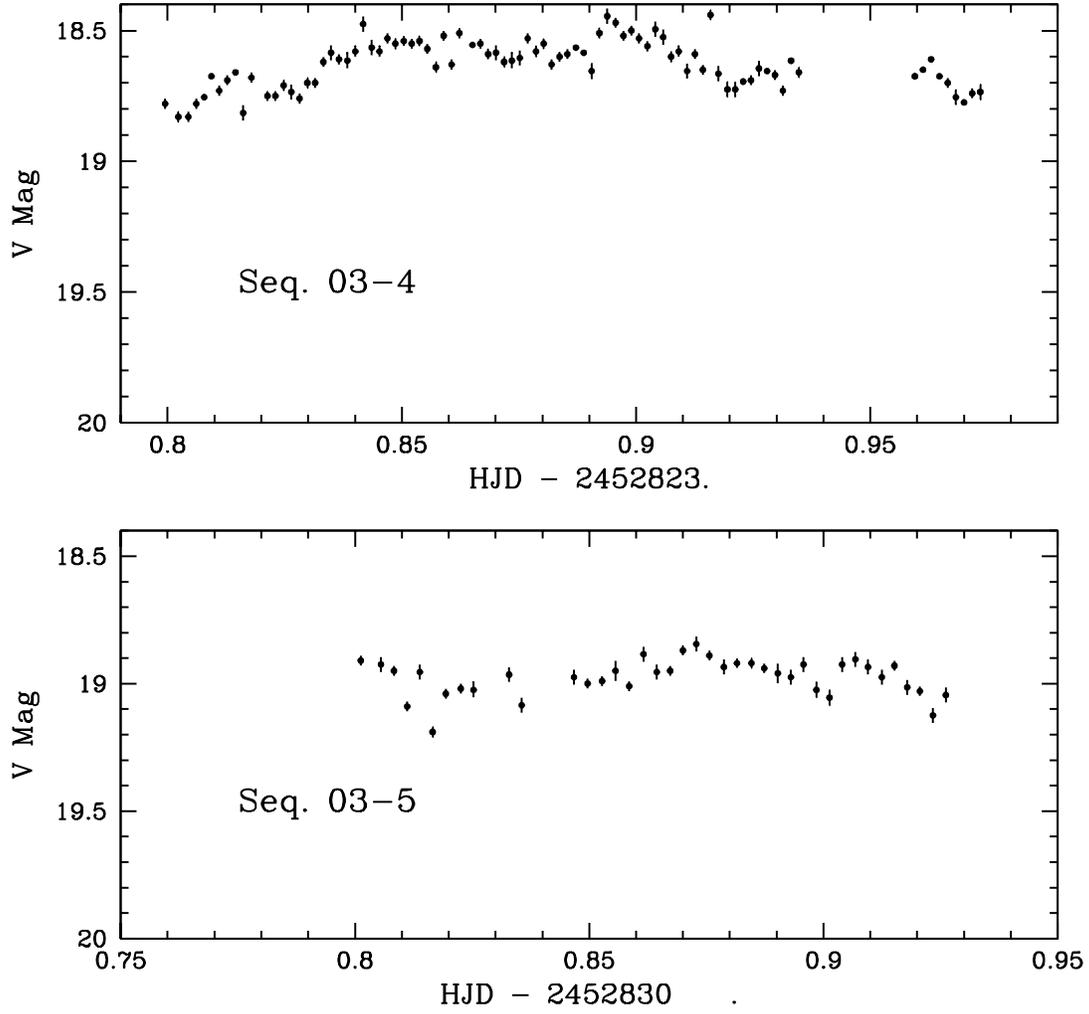}  
\caption{Mean magnitude of the low-state sequences for Set 1 (left) and for Set 2.  Sets
1 and 2 each encompass $\sim$ 3 weeks, separated by 1.5 years.  The short error bars
are for sdm, and the longer ones for sdso.} 
\end{figure}
\clearpage  

\begin{figure}  
\epsscale{0.9}
\plotone{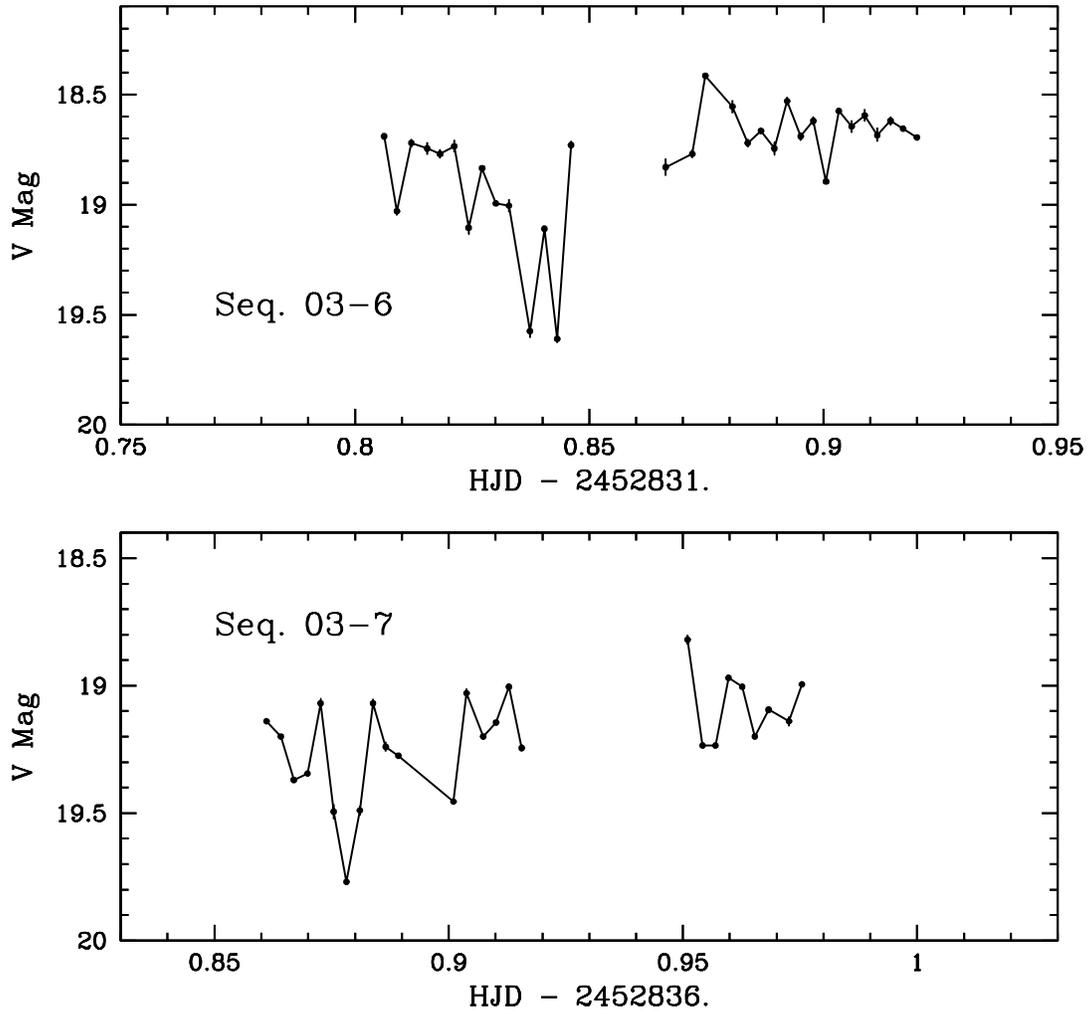}  
\caption{Average of 2 exposures of DK Lac obtained 2005-Oct-12 (UT) with the 
KPNO 2.1-m telescope (spectrum S1 as labeled in Figure 2 and Table 3).  The 
continuum has been normalized to
unity and the H$\alpha$ line profile is shown on an expanded scale.  No
corrections for atmospheric extinction have been applied; therefore the
telluric absorption bands remain.} 
\end{figure}
\clearpage  

\begin{figure}  
\epsscale{0.9}
\plotone{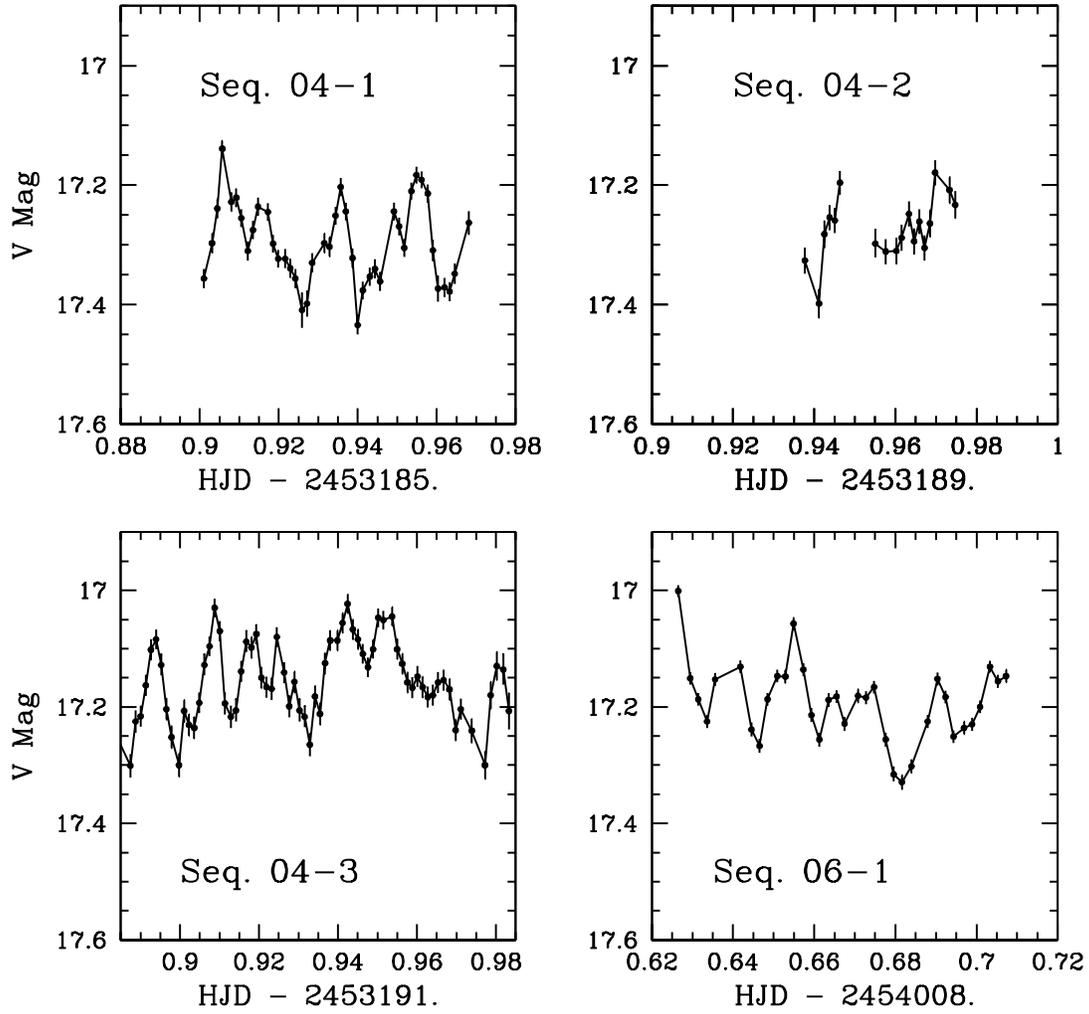}
\caption{Like Figure 7 except average of 3 high-state exposures of DK Lac 
2005-Oct-26 (UT) with the WIYN 3.5-m telescope (spectrum
S2 as labeled in Figure 2 and Table 3).}  
\end{figure}
\clearpage  

\begin{figure}  
\epsscale{0.9}
\plotone{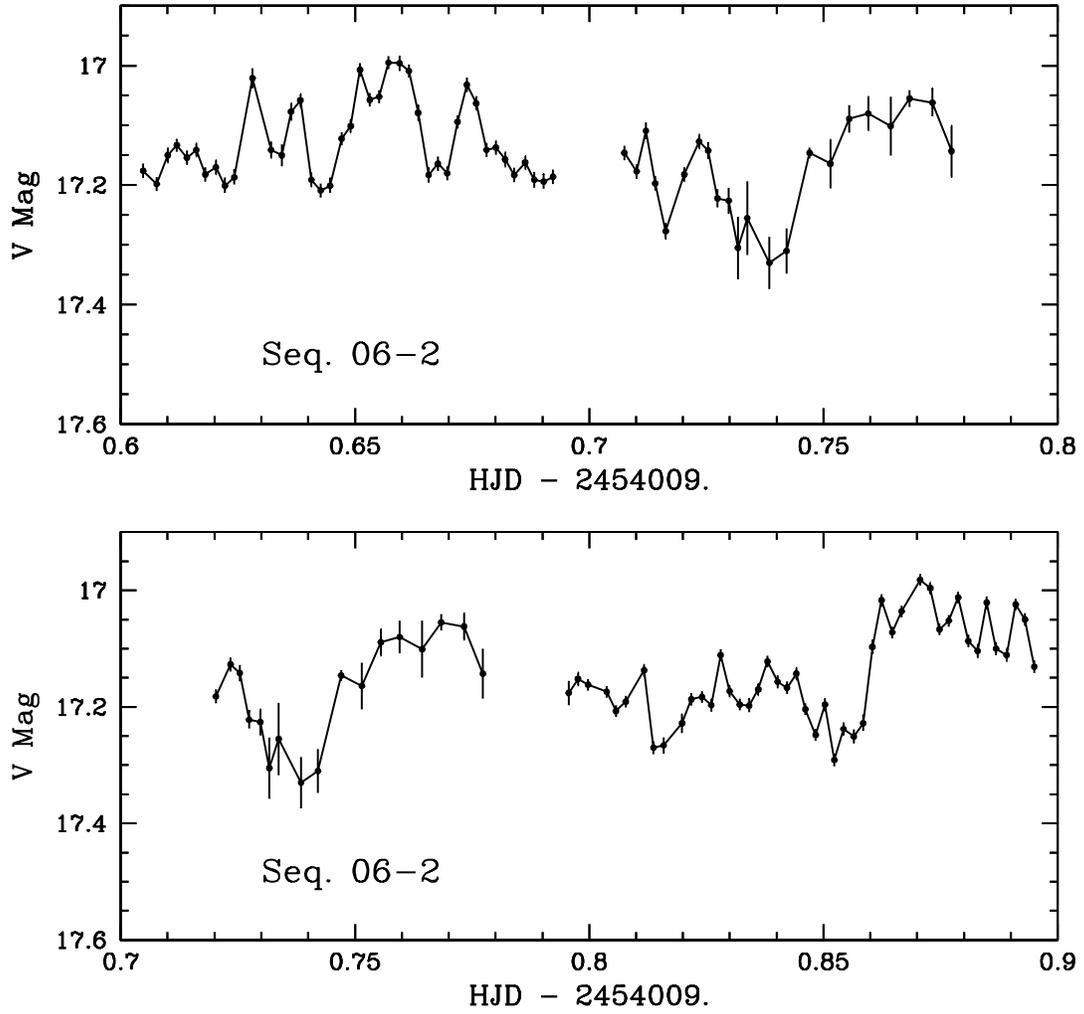}   
\caption{Like Figure 7 except average of 2 high-state exposures of DK Lac 
2006-Sep-16 (UT) with the MMT telescope (spectrum
S3 as labeled in Figure 2 and Table 3).} 
\end{figure}
\clearpage  

\begin{figure}  
\epsscale{0.9}
\plotone{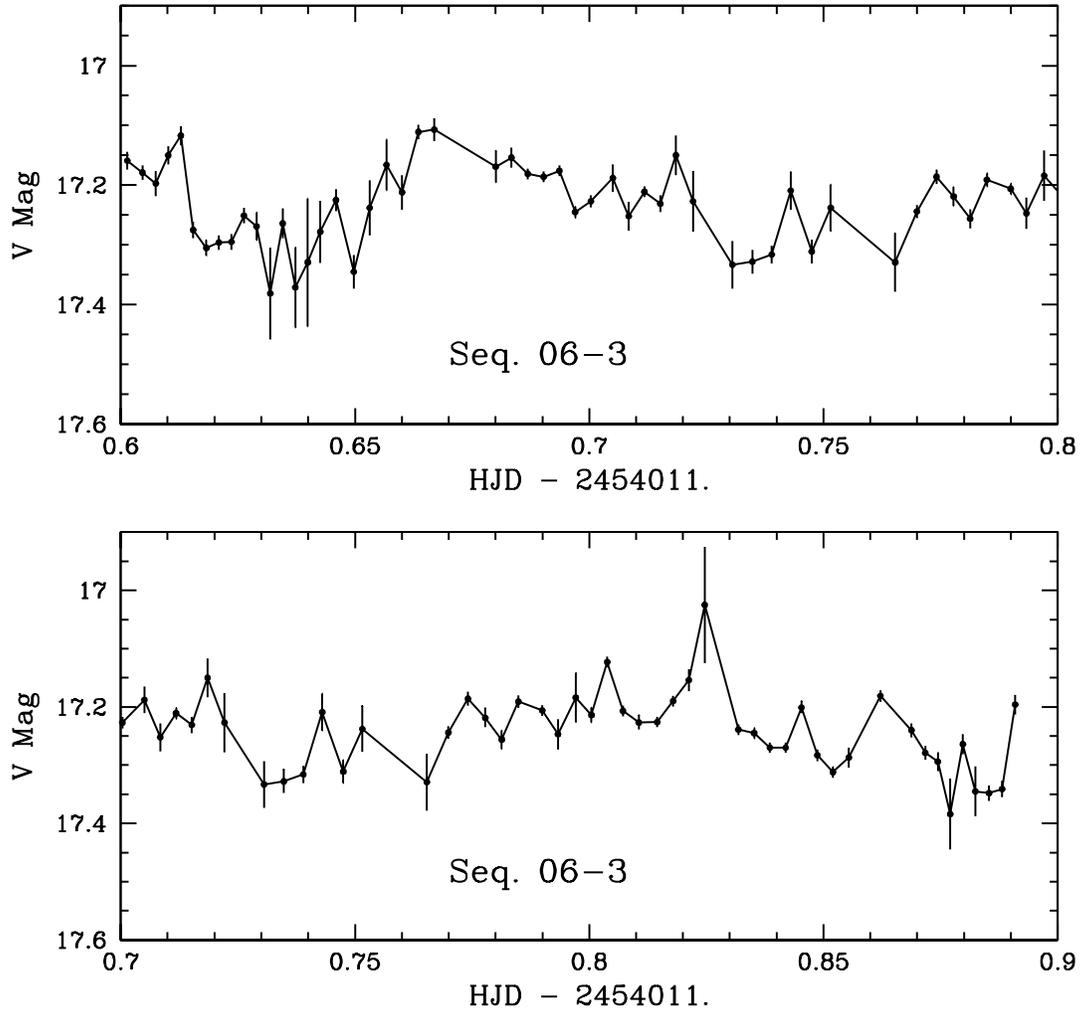}   
\caption{Like Figure 7 except average of 4 high-state exposures of DK Lac 
2008-Jun-07 (UT) with the KPNO 2.1-m telescope (spectrum
S4 as labeled in Figure 2 and Table 3).} 
\end{figure}
\clearpage  

\begin{figure}  
\epsscale{0.9}
\plotone{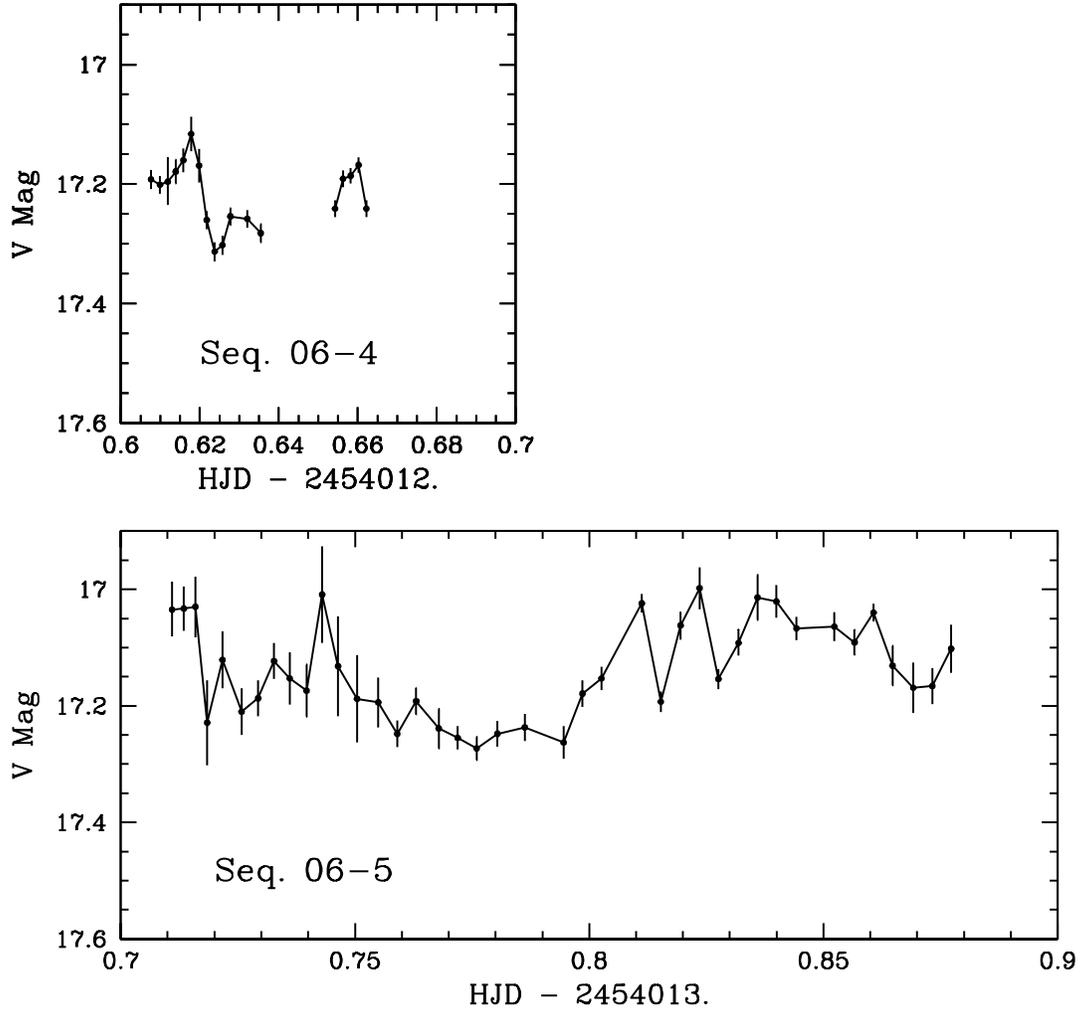}   
\caption{The equivalent width of H$\alpha$ in DK Lac as a function of JD, 1991-2008.  The
inset shows the most recent data points at sufficient scale to separate a closely-spaced
pair.} 
\end{figure}
\clearpage

\end{document}